\documentclass[aps,prb,reprint,twocolumn,superscriptaddress,showpacs,floatfix,longbibliography]{revtex4-1}

\usepackage{amsmath}
\usepackage{color}
\usepackage{mathrsfs}
\usepackage{dcolumn}
\usepackage{bm}
\usepackage{multirow}
\usepackage{graphicx}
\usepackage{rotating}
\usepackage{layout}
\usepackage{physics}
\usepackage[version=3]{mhchem}
\usepackage{braket}
\usepackage{longtable}
\usepackage{setspace}
\usepackage{booktabs}
\usepackage{siunitx}

\allowdisplaybreaks
\usepackage{xfrac}

\usepackage[colorlinks,citecolor=blue,urlcolor=blue,bookmarks=false,hypertexnames=true]{hyperref}

\begin{document}
\title{Self-consistent double-hybrid density functional theory via one-body second-order M{\o}ller-Plesset perturbation theory and projection-based embedding}
\date{\today}
\author{Huy Gia Bui}
\affiliation{Faculty of Physics and Engineering Physics, University of Science, Ho Chi Minh City 70000, Vietnam}
\affiliation{Vietnam National University, Ho Chi Minh City 70000, Vietnam}
\author{Lan Nguyen Tran}
\email{tnlan@hcmus.edu.vn}
\affiliation{Faculty of Physics and Engineering Physics, University of Science, Ho Chi Minh City 70000, Vietnam}
\affiliation{Vietnam National University, Ho Chi Minh City 70000, Vietnam}
\date{\today}

\begin{abstract}
We present the development of self-consistent one-body double-hybrid (OBDH) density functional theory (DFT). In this approach, the one-body second-order M{\o}ller-Plesset (OBMP2) perturbation potential is embedded directly in the generalized Kohn-Sham framework, allowing orbitals to be optimized in the presence of MP2-level dynamic correlation. Unlike existing orbital-optimized double hybrids, OBDH requires neither the optimized effective potential nor perturbative orbital relaxation corrections. The energy functional combines a semilocal exchange-correlation functional, exact exchange, and OBMP2 correlation, from which the effective Hamiltonian and self-consistent-field equations are systematically derived. To reduce computational cost, projector-based embedding with concentric localization truncation is applied to the OBMP2 component, termed sub-OBDH. OBDH and sub-OBDH are benchmarked on diatomic potential energy curves, self-interaction errors, dihedral torsions of organic molecules, and interaction energies in non-covalent charged systems. Across all systems, OBDH consistently outperforms standard DFT, demonstrating its potential for accurate and practical electronic structure calculations.
\end{abstract}
\maketitle

{\it Introduction.} 
Density functional theory (DFT), established on the foundational theorems of Hohenberg and Kohn~\cite{Hohenberg1964} and the practical self-consistent field framework of Kohn and Sham~\cite{Kohn1965}, has become the workhorse
of modern electronic structure calculations in chemistry and materials science. Its  balance between computational cost and accuracy has enabled the study of ground-state properties of systems ranging from small molecules to extended solids. Nevertheless, the predictive power of Kohn-Sham DFT is fundamentally contingent upon the quality of the exchange-correlation (XC) functional, whose exact form remains unknown and must be approximated in practice~\cite{Burke2012, Jones2015}.

Standard semilocal approximations, including the local density approximation (LDA)~\cite{Kohn1965, Perdew1992} and the generalized gradient approximation (GGA)~\cite{Perdew1996, Becke1988}, capture a broad range of physical and chemical phenomena, yet suffer from well-documented systematic deficiencies: self-interaction error~\cite{Perdew1981}, delocalization
error~\cite{Cohen2008}, and an inadequate description of London dispersion and long-range dynamical correlation effects~\cite{Klimes2012}.
These limitations motivate the continued development of more sophisticated XC approximations along the  Jacob's ladder of DFT proposed by Perdew and Schmidt~\cite{Perdew2001}.

A significant advance along this ladder was the introduction of hybrid functionals by Becke~\cite{Becke1993a}, which incorporate a fraction of nonlocal HF exchange into the XC functional. Within the generalized Kohn-Sham (GKS) framework formalized by Seidl, G\"orling, Vogl, Majewski, and Levy~\cite{Seidl1996}, the inclusion of nonlocal, orbital-dependent operators in the effective potential is rigorously justified. Hybrid functionals such as B3LYP~\cite{Becke1993a, Lee1988, Stephens1994}, PBE0~\cite{Adamo1999, Ernzerhof1999}, and HSE06~\cite{Heyd2003, Heyd2006} have demonstrated substantially improved performance for thermochemistry, reaction barriers, band gaps, and molecular geometries compared to their semilocal counterparts.

A further rung of Jacob's ladder is double-hybrid (DH) density functionals, originally proposed by Grimme~\cite{Grimme2006}, which augment the hybrid exchange with a perturbative second-order M\o{}ller-Plesset (MP2) correlation contribution~\cite{Moller1934}. Subsequent developments have produced a rich family of DH functionals, including B2-PLYP~\cite{Grimme2006},
mPW2-PLYP~\cite{Schwabe2006}, XYG3~\cite{Zhang2009}, PBE0-DH~\cite{Sharkas2011},
PBE-QIDH~\cite{Bremond2011}, and $\omega$B97X-2~\cite{Chai2009}, among many others~\cite{Goerigk2014, Sancho-Garcia2013, Bremond2016}. These functionals have demonstrated state-of-the-art accuracy across diverse benchmark sets including thermochemistry, kinetics, non-covalent interactions, and excited states~\cite{Goerigk2017, Mardirossian2017}.

Despite their success, conventional DH functionals suffer from a fundamental theoretical inconsistency. The orbitals entering the perturbative correlation expression are not variationally optimized with respect to the full DH energy functional\cite{Goerigk2014,DHFs-JCTC2020}. Therefore, the one-particle density matrix and related molecular properties are not fully consistent with the DH energy expression~\cite{Sharkas2011, Toulouse2011, Fromager2011}. Consequently, the development of a self-consistent framework for DH functionals remains an open challenge\cite{OODHFs-2013,OODHF-JPCA2016,OOMP2-OODHF-MP3-JCTC2020}.

One rigorous route to self-consistent inclusion of perturbative correlation within the KS framework is provided by the optimized effective potential (OEP) method~\cite{Sharp1953, Talman1976, Kummel2008}, which constructs a local multiplicative XC potential corresponding to an orbital-dependent functional. However, the OEP procedure is computationally demanding and numerically ill-conditioned in finite basis sets, limiting its practical applicability~\cite{Hirata2001, Bartlett2005, Grabowski2014}. Alternatively,
within the GKS framework, one may directly include the orbital-dependent correlation contribution through a nonlocal effective operator, bypassing the need for the OEP altogether~\cite{Seidl1996, Kummel2008}.

A particularly natural framework for wrapping MP2-level correlation self-consistently into a GKS orbital equation is offered by the one-body
M\o{}ller-Plesset second-order perturbation (OBMP2) theory, recently developed by one of the authors~\cite{OBMP2-JCP2013,OBMP2-JPCA2021,OBMP2-PCCP2022,OBMP2-JPCA2023,OBMP2-JPCA2024,OBMP2-JCP2025}. OBMP2 formulates the MP2 dynamic correlation effects as an effective one-body correlated Fock operator, obtained through a unitary canonical transformation of the molecular
Hamiltonian~\cite{CT-JCP2006,CT-JCP2007,CT-ACP2007,CT-JCP2009,CT-JCP2010,CT-IRPC2010} followed by the cumulant approximation to reduce many-body operators to one-body operators. Because OBMP2 yields a one-body correlated potential operator rather than a two-body perturbative energy, it is uniquely suited for direct incorporation into the GKS effective Hamiltonian without
recourse to response equations, orbital gradient corrections, or the OEP construction.

In this work, we present the development of self-consistent one-body double-hybrid (OBDH) DFT, in which the OBMP2 correlation potential is embedded in the GKS framework. The energy functional is constructed as a linear combination of semilocal XC functional, exact exchange (XX), and OBMP2 correlation. The OBDH effective Hamiltonian and associated SCF equations are derived using the GKS procedure\cite{Seidl1996,garrick2020exact}. In this approach, orbitals are self-consistently optimized in the presence of MP2-level dynamic correlation without invoking OEP or perturbative orbital relaxation corrections \cite{OODHFs-2013,OODHF-JPCA2016}. We further employ the Projector-based embedding (PbE) procedure \cite{PbE_2012,PbE_2019} in combination with Concentric Localization (CL) truncation\cite{CL_Claudino} for the OBMP2 component. OBDH is benchmarked on diatomic potential energy curves (PECs), self-interaction errors (SIEs), dihedral torsion of organic molecules, and interaction energy in non-covalent charged systems. These numerical results show that OBDH consistently improves upon standard DFT across these benchmarks.

{\it Theory.} The well-known GKS framework~\cite{Seidl1996, 
Kummel2008}  generalizes the standard Kohn--Sham (KS) formalism by allowing the model energy functional $\mathcal{S}[\{\phi_j\}]$ to depend explicitly on the orbitals rather than solely on the density. The universal Hohenberg--Kohn functional 
$F_\mathrm{HK}[n]$ is decomposed as
\begin{equation}
    F_\mathrm{HK}[n] = F_{\mathcal{S}}[n] + R_{\mathcal{S}}[n],
    \label{eq:HK_decomp}
\end{equation}
where $F_{\mathcal{S}}[n] \equiv \min_{\{\phi_j\}\to n(\mathbf{r})}\mathcal{S}[\{\phi_j\}]$ is 
the model energy and $R_{\mathcal{S}}[n]$ is the remainder energy functional. Minimizing the total energy with respect to the orbitals yields the GKS orbital equation:
\begin{equation}
    \Bigl[\hat{O}_{\mathcal{S}}[\{\phi_j\}] + V_\mathrm{ext}(\mathbf{r}) 
    + V_\mathrm{H}(\mathbf{r}) + V_{R}(\mathbf{r})\Bigr]\phi_i(\mathbf{r}) 
    = \varepsilon_i\,\phi_i(\mathbf{r}),
    \label{eq:GKS}
\end{equation}
where $V_{R}(\mathbf{r}) = \delta R_{\mathcal{S}}[n]/\delta n(\mathbf{r})$ is the local remainder 
potential and $\hat{O}_{\mathcal{S}}[\{\phi_j\}]$ is a generally nonlocal, non-multiplicative 
operator derived from $\mathcal{S}$. A central advantage of the GKS framework is that orbital-dependent contributions, such as exact exchange or correlated potentials, enter naturally through $\hat{O}_{\mathcal{S}}$ without invoking (OEP)~\cite{Sharp1953,Talman1976,Kummel2008,Hirata2001,Bartlett2005, Grabowski2014}.

The OBMP2 theory reformulates MP2-level dynamic correlation as an effective one-body operator, making it directly compatible with the GKS orbital equation~\cite{OBMP2-JCP2013,OBMP2-JPCA2021,OBMP2-PCCP2022,OBMP2-JPCA2023,OBMP2-JPCA2024,OBMP2-JCP2025}. Starting from a unitary canonical transformation~\cite{CT-JCP2006,CT-JCP2007,CT-ACP2007,CT-JCP2009,CT-JCP2010,CT-IRPC2010} of the molecular Hamiltonian $\hat{H}$ with an anti-Hermitian double-excitation operator $\hat{A} = -\hat{A}^\dagger$, and applying the cumulant approximation~\cite{cumulant-CPL1998,cumulant-JCP1997,cumulant-PRA1998,cumulant-JCP1999} to truncate many-body operators to one-body level, OBMP2 yields an effective Hamiltonian of the form
\begin{equation}
    \hat{H}_\mathrm{OBMP2} = \hat{H}_\mathrm{HF} + \hat{v}_\mathrm{OBMP2},
    \label{eq:HOBMP2}
\end{equation}
where $\hat{H}_\mathrm{HF} = \hat{F} + C$ is the HF Hamiltonian and $\hat{v}_\mathrm{OBMP2}$ 
is a one-body correlated potential operator. The full working expression for 
$\hat{v}_\mathrm{OBMP2}$ is given in Refs.~\cite{OBMP2-JCP2013,OBMP2-JPCA2021}. This can be written compactly as a correlated Fock operator,
\begin{equation}
    \hat{H}_\mathrm{OBMP2} = \bar{\hat{F}} + \bar{C},
    \label{eq:HOBMP2_compact}
\end{equation}
with the correlated Fock matrix
\begin{equation}
    \bar{f}^p_q = f^p_q + v^p_q,
    \label{eq:fbar}
\end{equation}
where $f^p_q$ is the standard HF Fock matrix and $v^p_q$ is the one-body correlation 
potential derived from the MP2 amplitudes
\begin{equation}
    T^{ab}_{ij} = \frac{g^{ab}_{ij}}{\varepsilon_i + \varepsilon_j - \varepsilon_a - \varepsilon_b}.
    \label{eq:T2}
\end{equation}
Here $\{i,j,\ldots\}$ and $\{a,b,\ldots\}$ denote occupied and virtual spin-orbital indices, respectively, and $g^{ab}_{ij}$ are the two-electron integrals. Because $\hat{v}_\mathrm{OBMP2}$ is a genuine one-body operator, it can be incorporated directly into a GKS effective Hamiltonian 
without recourse to response equations, orbital gradient corrections, or the OEP.

We now combine the GKS framework with OBMP2 theory to construct the self-consistent OBDH. The key idea is to define the GKS model functional $\mathcal{S}[\{\phi_j\}]$ as a linear 
combination of the noninteracting kinetic energy, the Hartree energy, a fraction $\alpha_x$ of exact HF exchange, and a fraction $\alpha_c$ of OBMP2 correlation:
\begin{equation}
\begin{aligned}
    \mathcal{S}[\{\phi_j\}] = &T_s[\{\phi_j\}]
    + \alpha_x\, E^\mathrm{HF}_x[\{\phi_j\}]
    + \alpha_c\, E^\mathrm{OBMP2}_c[\{\phi_j\}],
        \label{eq:OBDH_functional}
\end{aligned}
\end{equation}
where $T_s[\{\phi_j\}] = -\frac{1}{2}\sum_i\langle\phi_i|\nabla^2|\phi_i\rangle$ is the kinetic energy, $E^\mathrm{HF}_x[\{\phi_j\}]$ is the exact HF exchange energy, and $E^\mathrm{OBMP2}_c[\{\phi_j\}]$ is the OBMP2 correlation energy. 

Following the GKS procedure reported in Refs.~\citenum{Seidl1996} and \citenum{garrick2020exact}, we obtain the OBDH effective Hamiltonian:
\begin{equation}
\begin{aligned}    
    \hat{H}^\mathrm{OBDH}_\mathrm{eff} = &-\frac{1}{2}\nabla^2 + V_\mathrm{ext}(\mathbf{r}) 
    + V_\mathrm{H}(\mathbf{r}) + \alpha_x\,\hat{v}^\mathrm{HF}_x \\
    &+ V^\mathrm{DFA}_{xc}(\mathbf{r}) + \alpha_c\,\hat{v}_\mathrm{OBMP2},
    \label{eq:OBDH_Heff}
\end{aligned}
\end{equation}
where $\hat{v}^\mathrm{HF}_x$ is the nonlocal HF exchange operator, $V^\mathrm{DFA}_{xc} = (1-\alpha_x)V^\mathrm{DFA}_{x} + (1-\alpha_c)V^\mathrm{DFA}_{c}$ 
collects the semilocal DFA XC potential contributions, and $\hat{v}_\mathrm{OBMP2}$ is the one-body OBMP2 correlation operator from Eq.~\eqref{eq:HOBMP2}. Parameters $\alpha_x$ and $\alpha_c \in [0,1]$ control the mixing of exact exchange and OBMP2 correlation, respectively, recovering standard DFT at $\alpha_x = \alpha_c = 0$ and the full OBMP2 Hamiltonian in $\alpha_x = \alpha_c = 1$. In the current work, we use the Becke exchange potential\cite{Becke1993b} for $V^\mathrm{DFA}_{x}$ and Lee, Yang, and Parr (LYP) correlation potential\cite{LYP1988} for $V^\mathrm{DFA}_{c}$.

The resulting OBDH self-consistent field (SCF) equation reads
\begin{equation}
    \hat{H}^\mathrm{OBDH}_\mathrm{eff}\,\phi_i(\mathbf{r}) = \varepsilon_i\,\phi_i(\mathbf{r}).
    \label{eq:OBDH_SCF}
\end{equation}
The OBDH equation is solved self-consistently. At each SCF iteration, the amplitudes $T^{ab}_{ij}$ and the correlated potential $\hat{v}_\mathrm{OBMP2}$ are updated from the current orbitals and their eigenvalues. Convergence of this procedure yields orbitals that are optimized in the presence of both exact exchange and MP2-level dynamic correlation, with no recourse to the OEP, perturbative orbital relaxation, or response equations.

Since the formal scaling of OBMP2 is $\mathcal{O}(N^5)$, which may hinder OBDH from larger realistic applications, we incorporate the projector-based embedding (PbE) framework into OBDH to reduce the computational cost associated with the OB-MP2 component. PbE was developed by Miller, Manby and colleagues~\cite{PbE_2012,PbE_2019}, who demonstrated that the non-additive kinetic potential (NAKP) vanishes when the orbital sets of the active subsystem $A$ and the environment subsystem $B$ are mutually orthogonal, so that the kinetic energy of the total system reduces to the sum of the kinetic energies of the two subsystems. By enforcing this orthogonality via a level-shift projector $\mu P^B$ appended to the embedding Fock matrix, all difficulties associated with the non-additive kinetic energy (NAKE) completely disappear, and the total energy can be expressed in a clean, additive form. OBDH in the combination with PbE is denoted as sub-OBDH.

To partition the occupied orbital space into subsystems $A$ and $B$, we employ the Subsystem Projected AO DEcomposition (SPADE) procedure~\cite{SPADE_Claudino}, which performs a singular value decomposition (SVD) in the symmetrically orthogonalized basis to provide an automatic and unambiguous partition of the occupied space without affecting the virtual orbitals. In the limit of exact PbE, the environment orbitals of subsystem 
$B$ are shifted to arbitrarily high energies and contribute no correlation energy, leaving a reduced effective virtual space for the subsequent wave-function treatment. To further compress this virtual space, we apply the concentric localization (CL) protocol~\cite{CL_Claudino}, which projects the virtual orbitals onto a projection basis centered on the active atoms and then recursively generates localized virtual shells via successive SVDs of a 
one-particle operator. The active virtual space is built by retaining shells up to a chosen truncation level $n_{\text{shell}}$. The resulting localized virtual orbitals are finally pseudocanonicalized by diagonalizing the reference Fock matrix within the truncated subspace, rendering them fully compatible with the subsequent OB-MP2 treatment and significantly reducing its computational cost for large systems.

\begin{figure}[t!]
    \centering
    \includegraphics[width=1.0\linewidth]{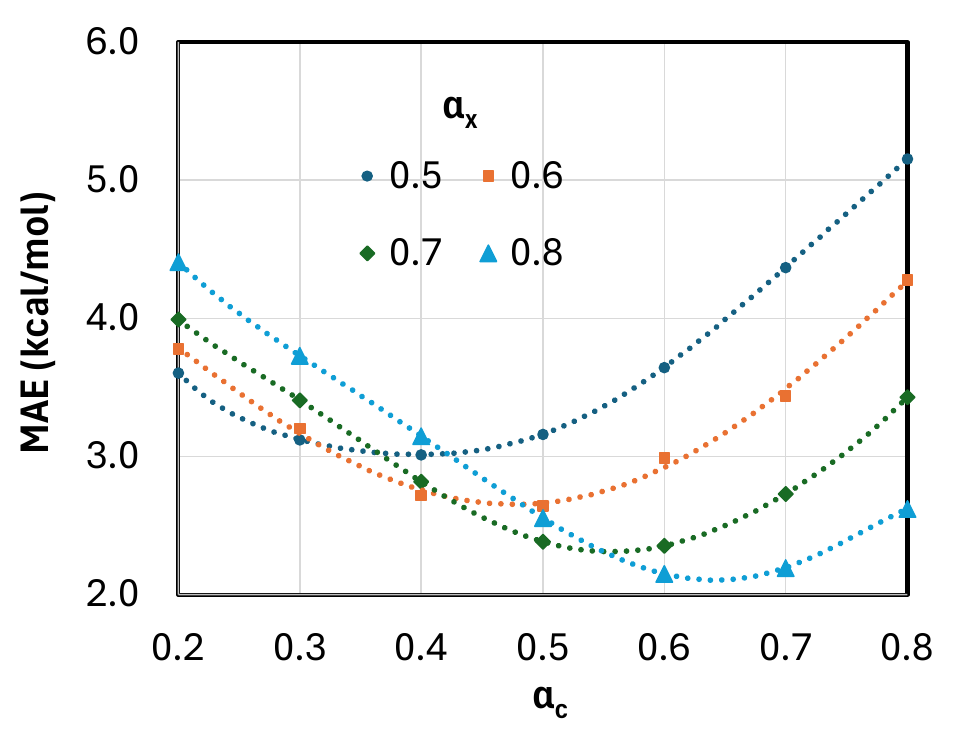}
    \caption{MAE (in kcal/mol) of ionization potentials calculated using various $\alpha_{x}$ and $\alpha_{c}$ scaling relative to the reference for the GMTKN55 database.}
    \label{fig:MAE_IPs}
\end{figure}

{\it Results.} We first seek optimal scaling factors $\alpha_x$ and $\alpha_c$. We note that the purpose here is not to find scaling factors that yield the smallest MAEs across many properties, but rather to identify a reasonable parametrization for further analysis of the advantages of OBDH in comparison with standard DFT and perturbation method. To this end, we employ the ionization potential (IP) set from the GMTKN55 database~\cite{GMTKN55_2017}. The MAEs of IPs calculated from OBDH with various scaling factors are shown in Figure~\ref{fig:MAE_IPs}. For each value of $\alpha_x$, there exists a corresponding value of $\alpha_c$ that minimizes the MAE. Based on this analysis, we select the pair $\alpha_x = 0.5$ and $\alpha_c = 0.4$, which yields an approximately balanced mixture of semilocal DFT, HF exchange, and OBMP2 correlation contributions.

\begin{figure*}[t!]
    \centering
    \includegraphics[width=0.8\linewidth]{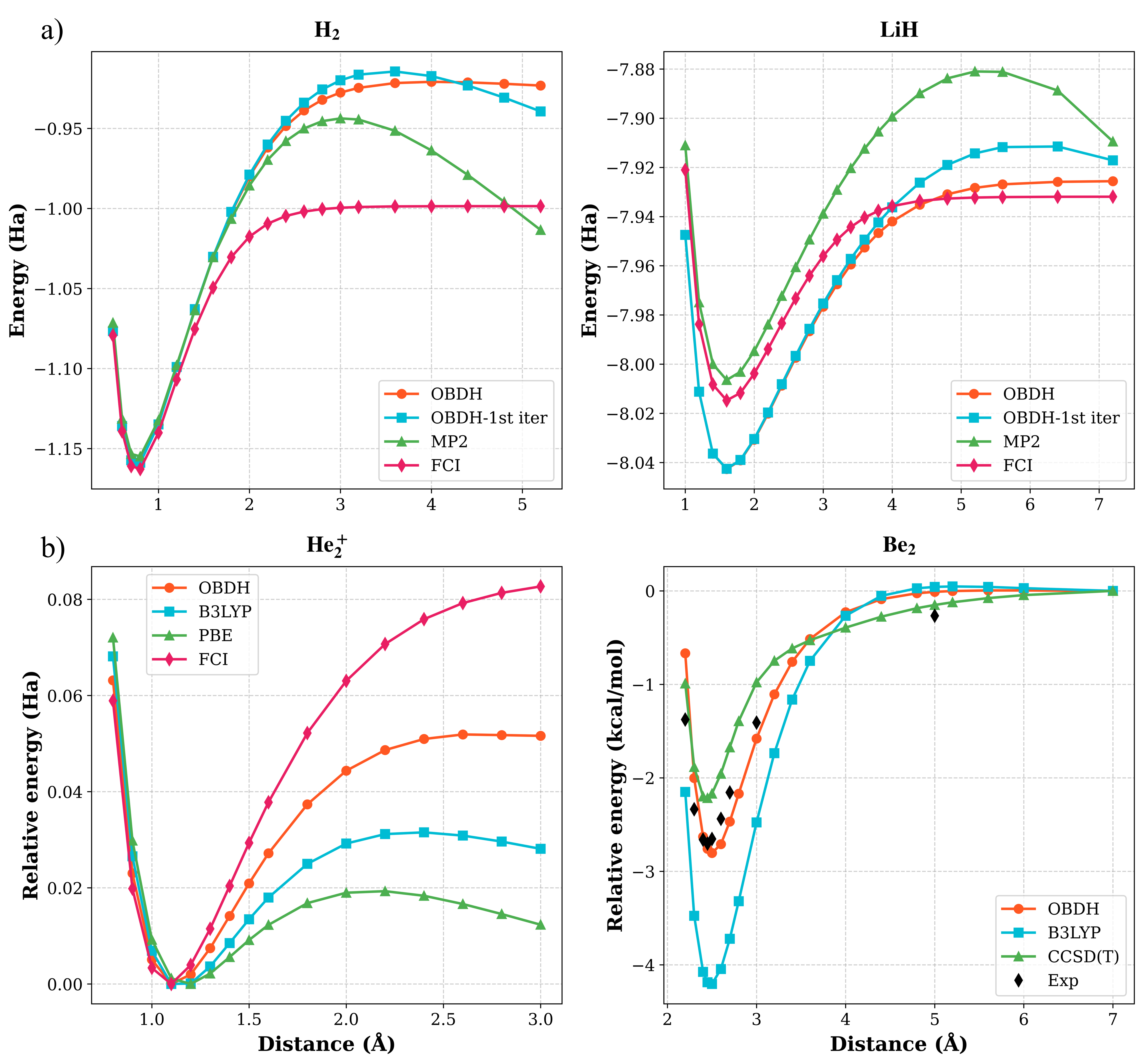}
    \caption{a) Potential energy curves of H$_2$ (left) and LiH (right) calculated using fully self-consistent OBDH, OBDH with the first iteration, MP2, and FCI. The cc-pVDZ basis set is used. b) Left: Potential energy curves of He$_2^+$ from OBDH, standard DFT (PBE and B3LYP), and FCI. Right: potential energy curves of Be$_2$ from OBDH, B3LYP, CCSD(T), and experiment. The cc-pVDZ and cc-pVQZ basis sets are used for He$_2^+$ and Be$_2$, respectively.}
    \label{fig:PES_H2LiH_He2+Be2}
\end{figure*}

One of the key drawbacks of non-iterative perturbation theory and standard double-hybrid functionals is that their dissociation curves diverge at long distances. To examine whether the self-consistent OBDH can overcome this issue, we plot in Figure~\ref{fig:PES_H2LiH_He2+Be2}a the potential energy curves of H$_2$ and LiH computed with OBDH, OBDH at the first iteration, MP2, and FCI in the cc-pVDZ basis. As is well known, although MP2 performs well near the equilibrium geometry, it diverges rapidly as the bond length increases for both systems. OBDH at the first iteration, which corresponds to a non-iterative double-hybrid functional, also diverges at long distances and fails to describe the dissociation correctly. Fully self-consistent OBDH, in contrast, describes the dissociation properly, with the potential energy curve running parallel to the FCI curve at long distances. Overall, the self-consistency in OBDH allows it to overcome the dissociation failures present in MP2 and non-iterative double-hybrid functionals.
   
Let us now examine OBDH on two systems that are challenging for standard DFT, He$_2^+$ and Be$_2$ as presented in Figure~\ref{fig:PES_H2LiH_He2+Be2}b. The raw data are given in the Supporting Information (SI). While cc-pVDZ is used for He$_2^+$, cc-pVQZ is used for Be$_2$ to compare to experiment. He$_2^+$ is a well-known example for testing DFT functionals against the self-interaction error (SIE) arising from the Coulomb energy terms that correspond to interaction of an electron with itself. As we can see in the left panel of Figure~\ref{fig:PES_H2LiH_He2+Be2}b, standard DFT functionals (PBE and B3LYP) yield reasonable results around equilibrium, but rapidly become too low in energy as distance increases, exhibiting unphysical barriers at dissociation limit. In contrast, OBDH does not display an energy barrier in the distance range considered here, implying that OBDH is able to reduce SIE present in standard DFT.

The ground state PEC of Be$_2$, which requires an accurate description of long-range dispersion interactions, is a well-known test for correlated theories \cite{Be2_2009}. Uncorrelated HF method yields a repulsive energy curve. While MP2 produces binding energies nearly three times too small compared to experiment, those from local and semi-local DFT are three to five times too large \cite{Be2_2005,Be2_2005_2}. RPA based on the PBE reference (RPA-PBE) produces too shallow a well depth and an unphysical repulsive barrier at intermediate bonding distances\cite{GKS_RPA}. Interestingly, while RPA-OEP does not remove the barrier and produces a positive well depth, GKS-spRPA not only removes the unphysical barrier but also considerably improves the well depth, yielding results close to the CCSD(T) ones~\cite{GKS_RPA}. In the right panel of Figure~\ref{fig:PES_H2LiH_He2+Be2}b, we plot the Be$_2$ PEC from OBDH, B3LYP, CCSD(T), and experiment. As can be seen, the B3LYP PEC exhibits binding energies that are too large, with a small unphysical barrier around 5.0 \r{A}. Although CCSD(T) yields a more physical PEC without an energy barrier, its binding energy is smaller than the experimental value. Interestingly, the OBDH PEC is closer to the experimental curve and does not produce an unphysical barrier observed in B3LYP.

\begin{table*}[t]
    \centering
    \caption{Self-interaction errors (SIE) of different functionals using the SIE4$\times$4 dataset\cite{GMTKN55_2017}. Signed errors for the computed reaction energies (in kcal/mol) relative to the reference adapted 
from Ref.~\citenum{GMTKN55_2017}, and mean absolute deviation (MAD) (kcal/mol) are reported.}
    \label{tab:sie4x4}
\begin{tabular}{llcrrrrr}
\hline \hline
\multirow{2}{*}{Reaction} 
    & \multirow{2}{*}{$R/R_{\mathrm{e}}$} 
    & \multirow{2}{*}{Reference\cite{GMTKN55_2017}} 
    & \multicolumn{4}{c}{Signed Error (kcal/mol)}\\
\cline{4-7}
 & & & OBDH & PBE & PBE0 & B3LYP \\
\hline
\multirow{4}{*}{\ce{H2+ -> H + H+}}
        & 1.00 & 64.4 & 26.3 & 54.8 & 40.2 & 44.7 \\
        & 1.25 & 58.9 & 24.5 & 50.9 & 37.4 & 41.5 \\
        & 1.50 & 48.7 & 22.5 & 46.7 & 34.5 & 38.1 \\
        & 1.75 & 38.3 & 20.5 & 42.3 & 31.4 & 34.5 \\[4pt]
\multirow{4}{*}{\ce{He2+ -> He + He+}}
        & 1.00 & 56.9 & 31.1 & 68.3 & 49.1 & 53.3 \\
        & 1.25 & 46.9 & 27.4 & 58.9 & 43.1 & 46.2 \\
        & 1.50 & 31.3 & 23.5 & 49.5 & 36.7 & 39.1 \\
        & 1.75 & 19.1 & 19.8 & 41.2 & 30.8 & 32.8 \\[4pt]
\multirow{4}{*}{\ce{(NH3)2+ -> NH3 + NH3+}}
        & 1.00 & 35.9 &  $-$3.8  & $-$11.3 &  $-$5.8 &  $-$6.2  \\
        & 1.25 & 25.9 &  $-$7.2  & $-$18.0 &  $-$9.4 & $-$11.7  \\
        & 1.50 & 13.4 & $-$10.4  & $-$24.4 & $-$13.4 & $-$16.6  \\
        & 1.75 &  4.9 & $-$13.2  & $-$29.8 & $-$17.1 & $-$20.7  \\[4pt]
\multirow{4}{*}{\ce{(H2O)2+ -> H2O + H2O+}}
        & 1.00 & 39.7 &  20.2 & $-$16.5 &  $-$8.0 &  $-$9.3  \\
        & 1.25 & 29.1 &  15.9 & $-$25.3 & $-$12.8 & $-$15.9  \\
        & 1.50 & 16.9 &  12.0 & $-$32.7 & $-$17.6 & $-$21.3  \\
        & 1.75 &  9.3 &   9.0 & $-$38.2 & $-$21.5 & $-$25.6  \\
\hline
\multicolumn{3}{l}{MAD (kcal/mol)} & 17.9 & 38.0 & 25.5 & 28.6 \\
\hline \hline
\end{tabular}
\end{table*}

We further analyze SIE using the SIE4x4 dataset~\cite{GMTKN55_2017}, which encompasses 16 dissociation reactions of cationic dimers at four internuclear separations ranging from the equilibrium geometry ($R/R_{\mathrm{e}} = 1.0$) to significantly stretched configurations ($R/R_{\mathrm{e}} = 1.75$). Errors relative to the reference~\citenum{GMTKN55_2017} from various DFT calculations are reported in Table~\ref{tab:sie4x4}. Raw data are given in SI.
PBE exhibits the worst performance across all systems, with errors growing systematically at stretched geometries, a well-known manifestation 
of SIE in pure GGA functionals. The hybrid functionals PBE0 and B3LYP partially alleviate this issue through the incorporation of Hartree-Fock exchange.
Interestingly, OBDH achieves the lowest mean absolute deviation (MAD) of 17.9~kcal/mol, representing an improvement over PBE (38.0~kcal/mol), B3LYP (28.6~kcal/mol), and PBE0 (25.5~kcal/mol), demonstrating that OBDH significantly mitigates 
the delocalization error inherent in conventional density functional approximations. 

\begin{figure*}[t!]
    \centering
    \includegraphics[width=0.75\linewidth]{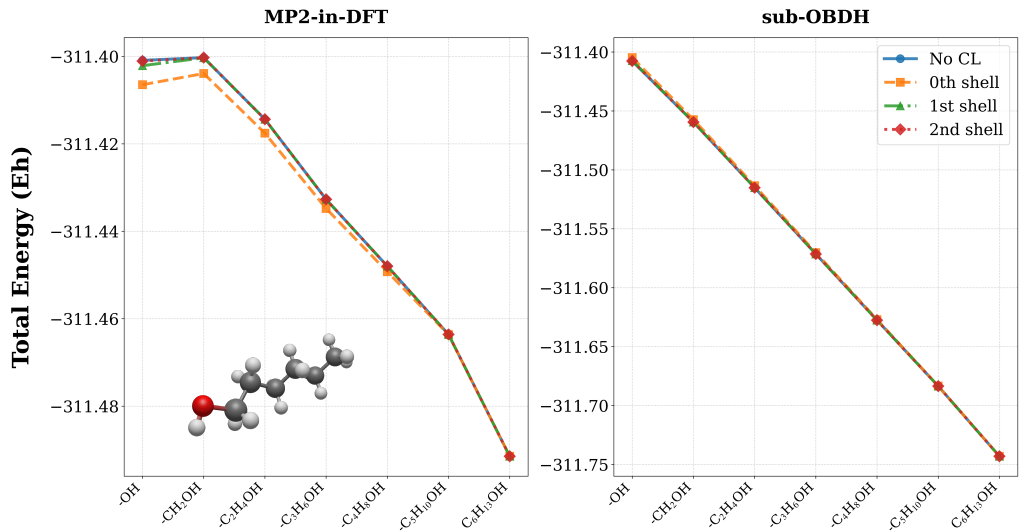}
    \caption{The change in energy when the embedded region is enlarged for CH$_3$(CH$_2$)$_5$OH.}
    \label{fig:hexanol}
\end{figure*}

We now turn to OBDH within the projector-based embedding framework, termed sub-OBDH. We first examine the convergence of sub-OBDH with respect to the size of the embedded region and the CL shell truncation. In Figure~\ref{fig:hexanol}, we plot the change in energy as the embedded fragment and CL truncation level increase for the CH$_3$(CH$_2$)$_5$OH (hexanol) molecule, with MP2-in-DFT shown for comparison. We note that, for consistency, we employ the same semilocal DFA and HF components in the environment for both sub-OBDH and MP2-in-DFT. We can see that, while the sub-OBDH energy monotonically decreases as the embedded region is enlarged, an unphysical increase of energy is observed from --OH to --CH$_2$OH in the MP2-in-DFT results. Furthermore, the energy decrease of sub-OBDH is more linear than that of MP2-in-DFT.
This linear convergence behavior is physically consistent with the electronic structure of hexanol, which consists of $\sigma$-bonded C--C and C--O single bonds with no $\pi$-conjugation or long-range electronic delocalization. In such systems, each additional methylene (--CH$_2$--) unit incorporated into 
the embedded region contributes an approximately equal and independent correction to the high-level energy, resulting in a uniform, additive decrease of the embedding error with each successive shell. 

Regarding the effect of CL shell truncation, the two methods exhibit markedly different sensitivities. For sub-OBDH, the curves corresponding to different CL shell levels ($0^{\text{th}}$, $1^{\text{st}}$, and $2^{\text{nd}}$ shells) overlap almost perfectly throughout the entire range of embedded region sizes, indicating that the sub-OBDH energy is 
insensitive to the extent of CL truncation. This rapid convergence with respect to the CL shell suggests that the correlation contributions in sub-OBDH are well-localized and that a minimal CL description is already sufficient to capture the essential physics. In contrast, MP2-in-DFT displays a more pronounced discrepancy among the different CL shell levels, particularly in the smaller embedded region regime, where the differences between the $0^{\text{th}}$ shell and others  are clearly visible. 
The faster convergence of sub-OBDH with respect to CL truncation thus represents a practical advantage, as it reduces the computational cost associated with constructing large CL shells while maintaining accuracy.

\begin{figure*}[t!]
    \centering
    \includegraphics[width=0.8\linewidth]{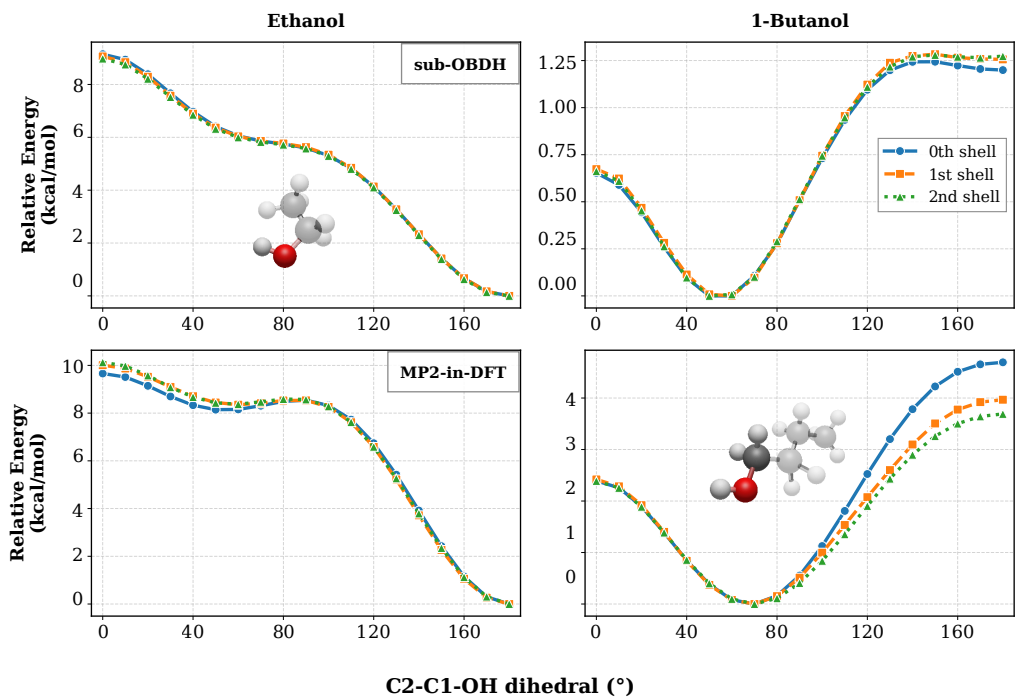}
    \caption{Relative energy profiles as a function of the C2--C1--O--H dihedral angle for ethanol (left) and 1-butanol (right), computed using sub-OBDH (top) and MP2-in-DFT (bottom) with different CL shell truncation levels ($0^{\text{th}}$, $1^{\text{st}}$, and $2^{\text{nd}}$ shells). The embedded regions for Ethanol and 1-Butanol are --OH and --CH$_2$OH, respectively.}
    \label{fig:torsion}
\end{figure*}

To further assess the convergence with respect to CL shell truncation, we examine the relative energy profiles along the C2--C1--O--H dihedral angle for ethanol and 1-butanol in Figure~\ref{fig:torsion}. Raw data are given in SI. These torsional profiles provide a more stringent test of the embedding quality, as they probe the sensitivity of the method to the chemical environment beyond the embedded --OH fragment across a range of molecular geometries. For sub-OBDH, the curves corresponding to the $0^{\text{th}}$, $1^{\text{st}}$, and $2^{\text{nd}}$ CL shells are very close to each other for both ethanol and 1-butanol throughout the entire dihedral scan. This insensitivity to CL truncation level confirms that sub-OBDH achieves rapid convergence with respect to the size of the environment, even for the larger 1-butanol system where long-range interactions might be expected to play a more significant role. The result suggests that the correlation contributions captured by sub-OBDH are inherently local and do not require an extensive CL description to be accurately represented. 

In stark contrast, MP2-in-DFT exhibits a strong dependence on the CL shell truncation level, particularly for 1-butanol. The $0^{\text{th}}$, $1^{\text{st}}$, and $2^{\text{nd}}$ shell curves show visible discrepancies, especially in the high-energy region of the dihedral profile (around 90°--120°), where the separation between curves is most pronounced. This sensitivity indicates that MP2-in-DFT requires a more complete description of the environment to achieve convergence, reflecting its greater dependence on the CL region to account for the electronic influence of atoms outside the embedded fragment. The fact that these discrepancies grow with molecular size, from ethanol to 1-butanol, further highlights the limitation of MP2-in-DFT in terms of CL convergence for larger systems. In general, the results in Figures~\ref{fig:hexanol} and \ref{fig:torsion} consistently demonstrate that sub-OBDH converges more rapidly and reliably with respect to CL shell truncation than MP2-in-DFT, offering a more robust and computationally efficient embedding framework.

\begin{figure*}[t!]
    \centering
    \includegraphics[width=1.0\linewidth]{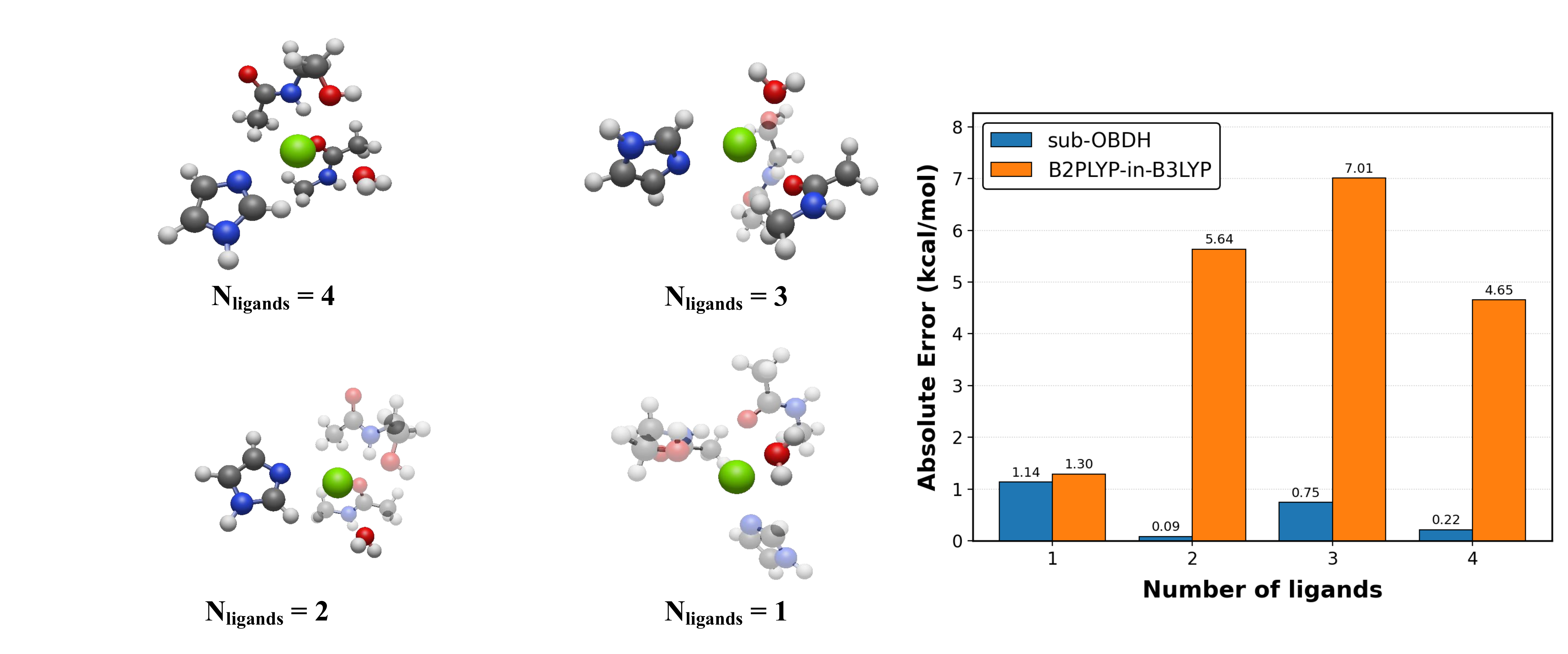}
    \caption{Left panel: Subcluster generation for the Mg$^{2+}$ cluster (PDB ID 2YNV), sequential ligand removal from the original cluster yielding subclusters with one to five coordinating ligands. Right panel: Absolute interaction energy errors (Abs. err.) of sub-OBDH and B2PLYP-in-B3LYP relative to the LNO-CCSD(T) reference\cite{NIC_charged_26} for each system in the left panel. 
    }
    \label{fig:nci}
\end{figure*}

Noncovalent interactions (NCIs) involving charged systems are ubiquitous in 
biochemistry, catalysis, and materials science, playing essential roles in 
enzymatic catalysis, protein folding, and ion transport~\cite{NIC_charged_13,NIC_charged_19,NIC_charged_23}. Despite its widespread use, standard dispersion-enhanced DFT exhibits large errors for charged NCIs\cite{NIC_DFT_16,NIC_DFT_23}, primarily due to delocalization 
errors and the inherent coupling between electrostatics, polarization, and dispersion in inhomogeneous electric fields.
To address these limitations, Zhao et al.~\cite{NIC_charged_26} introduced (r$^2$SCAN+MBD)@HF, a parameter-free composite approach that evaluates the r$^2$SCAN meta-GGA functional and many-body dispersion (MBD) on HF orbitals, thereby correcting density-driven errors while ensuring balanced short- and long-range correlation. In contrast, PBE0+MBD is a conventional hybrid DFT approach that, while broadly used, suffers from overbinding in charged 
systems due to inadequate treatment of polarization and dispersion coupling.

Let us consider the Mg$^{2+}$ cluster (PDB ID 2YNV), depicted on the left panel of Figure~\ref{fig:nci}. For comparison, we also performed the B2PLYP-in-B3LYP embedding. The basis set def2-TZVPD was used for all calculations to be consistent with the previous work by Zhao et al.~\cite{NIC_charged_26}.
While the $N_\mathrm{ligands} = 1$ cluster is treated in full, without embedding, active regions are assigned for larger systems by atom count as follows: Mg$^{2+}$ and \ce{H2O} for $N_\mathrm{ligands} = 2$ and $3$ (4 of 13 and 4 of 25 atoms, respectively), and Mg$^{2+}$, \ce{H2O}, and imidazole for $N_\mathrm{ligands} = 4$ (13 of 41 atoms). The CL truncation was employed with $n_{\text{shell}} = 2$ for both sub-OBDH and B2PLYP-in-B3LYP.
Figure~\ref{fig:nci} presents the absolute errors of sub-OBDH and B2PLYP-in-B3LYP interaction energies relative to LNO-CCSD(T) for a series of complexes with increasing the number of ligands. Raw data are given in SI. As reported in Ref~\citenum{NIC_charged_26}, PBE0+MBD errors are dramatically increased. While (r$^2$SCAN+MBD)@HF yields much smaller errors than PBE0+MBD, it also exhibits a monotonic increase of errors. Interestingly, we can see here that the sub-OBDH maintains more stable errors across all systems, reflecting its robustness. While B2PLYP-in-B3LYP has errors up to $\sim7.0$ kcal/mol for $N_\mathrm{ligands} = 3$, sub-OBDH yields much smaller errors with the largest one being only $\sim1.1$ kcal/mol, implying that it is highly promising for noncovalent interactions in charged systems. 

\textit{Conclusion.}
We have presented OBDH, a self-consistent double-hybrid density functional framework built on the GKS formalism and the OBMP2 theory. The central idea is that the one-body operator structure of OBMP2 allows second-order dynamic correlation to be embedded directly and self-consistently into the GKS effective Hamiltonian, bypassing the need for OEP construction or perturbative orbital relaxation corrections. The resulting OBDH orbital equation variationally optimizes the orbitals in the presence of both exact exchange and MP2-level correlation on equal footing.

Numerical results confirm the theoretical advantages of the self-consistent treatment. For bond dissociation, fully self-consistent OBDH correctly describes the potential energy curves of H$_2$ and LiH, whereas non-iterative double-hybrid functionals and MP2 both diverge at stretched geometries. For SIE, OBDH achieves MAD of 17.9~kcal/mol on the SIE4$\times$4 dataset, a noticeable improvement over PBE, PBE0, and B3LYP. OBDH also correctly reproduces the weakly bound potential energy curve of Be$_2$, a known challenge for both DFT and low-order perturbation theory.

To extend OBDH to larger systems, we introduced sub-OBDH, which combines 
OBDH with the PbE procedure and the CL truncation. Sub-OBDH converges markedly faster with respect to CL shell truncation than MP2-in-DFT, as demonstrated for torsional profiles of ethanol and 1-butanol and the 
energy convergence of CH$_3$(CH$_2$)$_5$OH. 
For noncovalent interactions in metal-ligand complexes, sub-OBDH is substantially more accurate than B2PLYP-in-B3LYP, maintaining stable errors below $\sim$1.1 kcal/mol across systems of increasing size. It avoids the systematic deterioration observed in PBE0+MBD \cite{NIC_charged_26} and offers a unified, self-consistent alternative to composite schemes, such as (r$^2$SCAN+MBD)@HF. \cite{NIC_charged_26}

In general, numerical results establish OBDH and sub-OBDH as a theoretically well-grounded and practically competitive route to 
fully self-consistent double-hybrid DFT. A comprehensive benchmark across thermochemistry, kinetics, and molecular properties is currently underway.

\section*{Acknowledgments}
This research is funded by Vietnam National University,  Ho Chi Minh City (VNU-HCM) under grant number B2026-18-18. We thank Nhi Vo for initial work on the OBDH project. 

\bibliography{main}

\end{document}